\documentclass[twoside]{amsart}
\usepackage{latexsym}
\usepackage{amssymb,amsmath,amsopn,setspace}
\usepackage[dvips]{graphicx}   
\usepackage{color,epsfig}      

\newcommand{\mathsym}[1]{{}}
\newcommand{\unicode}[1]{{}}

\newtheorem{theorem}{Theorem}[]

\theoremstyle{definition}
\newtheorem{defi}[]{Definition}[]
\newtheorem{remark}[]{Remark}[]

\numberwithin{equation}{section}

\def\arc{\mathrm { arc }}

\begin{document}

\title{Relativity theory in time-space}
\author{\'Akos G.Horv\'ath}
\address{Department of Geometry, Mathematical Institute \\ Budapest University of Technology and Economics }
\email{ghorvath@math.bme.hu}
 %

\begin{abstract}
The concept of time-space defined in an earlier paper of the author is a certain generalization of the so-called space-time. In
this paper we introduce the concept of time-space manifolds. In the homogeneous case, a time-space manifold is a differentiable manifold with
such tangent spaces which have a certain fixed time-space structure. We redefine the fundamental concepts of global relativity theory with respect to this general situation. We study the concepts of affine connection, parallel transport, curvature tensor and Einstein equation, respectively.
\end{abstract}

\maketitle

\newtheorem{axiom}{Axiom}
\def\arc{\mathrm { arc }}

\section*{Introduction}

In \cite{gho1} we constructed a model on the basis of two types of Minkowski spaces, the space with indefinite inner product
(\emph{Lorentzian-Minkowski space} see e.g. \cite{gohberg}, \cite{minkowski}) and the space with a semi-inner product (finite-dimensional
separable Banach space see in \cite{giles}, \cite{lumer}, \cite{martini-swanepoel 1} and \cite{martini-swanepoel 2}). Among other things,
we introduced the concept of \emph{generalized Minkowski space} and especially the so-called \emph{generalized space-time model}, which is a
generalization of the Minkowski-Lorentz space-time. From differential geometric point of view, we investigated the latter in \cite{gho2}. This
investigation led to a generalization of the spaces of constant curvature; the hyperbolic (anti-de Sitter), de Sitter, and Euclidean spaces,
respectively. In its own right, in a generalized space-time there is a theory of special relativity which was not developed in the above
mentioned theoretical papers. In \cite{gho4} the concept of generalized space-time model was extracted to a model called \emph{generalized
Minkowski space with changing shape} (briefly \emph{time-space}). We gave two types of models, a non-deterministic (random) variation and a
deterministic one. We proved that in a finite range of time the random model can be approximated in an appropriate deterministic
model. Thus, from practical point of view the deterministic models are more important. We must mention here that the measure of a random model
is based on the following observation: on the space of norms such a geometric measure can be defined in such a way that its push-forward onto the line of the absolute-time has normal distribution (see \cite{gho3}).

A time-space can be given also via the help of the so-called \emph{shape function}. In Section 2 we give the fundamental formulas of special
relativity in a time-space (depending on the given shape function). In Section 3 we embed some known metrics of general relativity into a
suitable time-space. This shows that time-space is a good place to visualize some of these. Of course, since time-space has a direct product
character hence a lot of metrics holding the Einstein's equation have no natural embedding into it. In the last subsection of Section 3 we
define a generalization of the Lorentzian manifold which we call \emph{time-space manifold}. The tangent spaces of a time-space manifold are
time-spaces with linear shape-functions. We introduce the concept of \emph{homogeneous time-space manifold} as such a time-space manifold whose
tangent spaces can be identified with the same time-space. In a homogeneous time-space manifold we define concepts of global relativity
theory: affine connection, parallel transport, curvature tensor and Einstein equation.

The first paragraph contains those definitions, notations and statements which are used in this paper.

\subsection*{Deterministic and random time-space models}

We assume that there is an absolute coordinate system of dimension $n$ in which we are modeling the universe by a time-space model. The origin
is a generalized space-time model (see in \cite{gho1}) in which the time axis plays the role of absolute time.  Its points are unattainable
and immeasurable for me and the corresponding line is also in the exterior of the modeled universe. (We note that in Minkowskian space-time this
assumption holds only for the axes determining the space-coordinates.) This means that in our model, even though the axis of time belongs to the
double cone of time-like points, its points do not belong to the modeled universe. at a fixed moment of time (with respect to this absolute
time) the collection of the points of space can be regarded as an open ball of the embedding normed space centered at the origin and does
not contain the origin. The omitted point is the origin of a coordinate system giving the space-like coordinates of the world-points with
respect to our time-space system. Since the points of the axis of absolute-time are not in our universe there is no reference system in our
modeled world which determines absolute time.

In our probabilistic model (based on a generalized space-time model) the absolute coordinates of points are calculated by a fixed basis of the
embedding vector space. The vector $s(\tau)$ means the collection of the space-components with respect to the absolute time $\tau$, the quantity
$\tau$ has to be measured on a line $T$ which is orthogonal to the linear subspace $S$ of the vectors $s(\tau)$. (The orthogonality was
considered as the Pythagorean orthogonality of the embedding normed space.) Consider a fixed Euclidean vector space with unit ball $B_E$ on $S$
and use its usual functions e.g.  volume, diameter, width, thinness and Hausdorff distance, respectively. With respect to the moment $\tau$ of
the absolute time we have a unit ball $K(\tau)$ in the corresponding normed space  $\{S,\|\cdot\|^{\tau}\}$. The modeled universe at $\tau$ is
the ball $\tau K(\tau)\subset \{S,\|\cdot\|^{\tau}\}$. The shape of the model at the moment $\tau$ depends on the shape of the centrally
symmetric convex body $K(\tau)$. The center of the model is on the axis of absolute time, it cannot be determined. For calculations on
time-space we need further smoothness properties on the function $K(\tau)$. These are
\begin{itemize}
\item $K(\tau)$ is a centrally symmetric,
convex, compact, $C^2$-class body of volume $\mathrm{vol}(B_E)$. \item For each pairs of points $s',s''$ the function $$ K:\mathbb{R}^+\cup
\{0\}\rightarrow \mathcal{K}_0 \mbox{ , }\tau\mapsto K(\tau) $$ holds the property that $[s',s'']^{\tau}:\tau\mapsto [s',s'']^{\tau}$ is a
$C^1$-function.
\end{itemize}

\begin{defi}
We say that a generalized space-time model endowed with a function $K(\tau)$ holding the above properties is a \emph{deterministic
time-space model}.
\end{defi}

The main subset of a deterministic time-space model contains the points of negative norm-square. This is the set of time-like points and the
upper connected sheet of the time-like points is the modeled universe. The points of the universe have positive time-components. We denote this
model by $ \left(M,K(\tau)\right). $

To define a random time-space model we should choose the function $K(\tau)$ ``randomly". To this purpose we use Kolmogorov's extension theorem
(or theorem on consistency, see \cite{kolmogorov}). This says that a suitably "consistent" collection of finite-dimensional distributions
will define a probability measure on the product space. The sample space here is $\mathcal{K}_0$ with Hausdorff distance. It is a locally
compact, separable (second-countable) metric space. By Blaschke's selection theorem $\mathcal{K}$ is a boundedly compact space so it is also
complete. It is easy to check that $\mathcal{K}_0$ is also a complete metric space if we assume that the non-proper bodies (centrally symmetric
convex compact sets with empty interior) also belong to it. (In the remaining part we regard such a body as the unit ball of a normed space of
smaller dimension.) Finally, let $P$ be a probability measure. At every moment of absolute time we consider the same probability space
$\left(\mathcal{K}_0, P\right)$ and also consider in each of the finite collections of moments the corresponding product spaces
$\left((\mathcal{K}_0)^r, P^r\right)$ . The consistency assumption of Kolmogorov's theorem now automatically holds. By the extension theorem we
have a probability measure $\hat{P}$ on the measure space of the functions on $T$ to $\mathcal{K}_0$ with the $\sigma$-algebra generated by the
cylinder sets of the space. The distribution of the projection of $\hat{P}$ to the probability space of a fix moment is the distribution of
$P$.

\begin{defi} Let $(K_\tau \mbox{ , }\tau\geq 0)$ be a random function defined as an element of the Kolmogorov's extension $\left(\Pi
\mathcal{K}_0, \hat{P}\right)$ of the probability space $\left(\mathcal{K}_0, P\right)$. We say that the generalized space-time model with the
random function $$ \hat{K}_\tau:=\sqrt[n]{\frac{\mathrm{ vol}(B_E)}{\mathrm{ vol}(K_\tau)}}K_\tau $$ is a \emph{random time-space model}. Here
$\alpha_0(K_\tau)$ is a random variable with truncated normal distribution and thus $(\alpha_0(K_\tau) \mbox{ , } \tau\geq 0)$ is a stationary
Gaussian process. We call it the \emph{shape process} of the random time-space model.
\end{defi}
It is clear that a deterministic time-space model is a special trajectory of the random time-space model. The following theorem is essential.
\begin{theorem}[\cite{gho4}]
For a trajectory
$L(\tau)$ of the random time-space model, for a finite set $0\leq \tau_1\leq \cdots \leq \tau_s$  of moments and for $\varepsilon >0$ there
is a deterministic time-space model defined by the function $K(\tau)$ for which
$$
\sup\limits_{i}\{\rho_H\left(L(\tau_i), K(\tau_i)\right)\}
\leq \varepsilon.
$$
\end{theorem}
An important consequence of Theorem 1 is the following: \emph{ Without loss of generality we can assume that the time-space model is deterministic.}

\begin{defi} For two vectors $s_1+\tau_1$ and $s_2+\tau_2$ of the deterministic time-space model we define their product with the equality
$$
[s_1+\tau_1,s_2+\tau_2]^{+,T}:=[s_1,s_2]^{\tau_2}+\left[\tau_1,\tau_2\right]=
$$
$$
=[s_1,s_2]^{\tau_2}-\tau_1\tau_2.
$$
\end{defi}
Here
$[s_1,s_2]^{\tau_2}$ means the s.i.p defined by the norm $\|\cdot\|^{\tau_2}$. This product is not a Minkowski product, as there is no
homogeneity property in the second variable. On the other hand, the additivity and homogeneity properties of the first variable, the properties
on non-degeneracy of the product again hold.  Finally, the continuity and differentiability properties of this product also
remain the same as of a Minkowski product. The calculations in a generalized space-time model basically depend on a rule on the
differentiability of the second variable of the Minkowski product. As a basic tool of investigations we proved in \cite{gho4} that

\begin{theorem}[\cite{gho4}] If $f_1, f_2: S\longrightarrow V=S+T$ are two $C^2$ maps and $c:\mathbb{R}\longrightarrow S$ is an arbitrary $C^2$
curve then
$$ ([(f_1\circ c)(t)),(f_2\circ c)(t))]^{+,T})'=
$$
$$ =[D(f_1\circ c)(t),f_2(c(t))]^{+,T}
+\left({[f_1(c(t)),\cdot]^{+,T}}\right)'_{D(f_2\circ c)(t)}(f_2(c(t)))+
$$
$$
+\frac{\partial{\left[(f_1)_S(c(t)),(f_2)_S(c(t))
\right]^{\tau}}}{\partial \tau}((f_2)_T(c(t)))\cdot((f_2)_T\circ c)'(t).
$$
\end{theorem}

The theory of generalized space-time model can be used in a generalization of special relativity theory, if we change some previous formulas
using also the constant $c$. ($c$ can practically be considered as the speed of light in vacuum.) The formula of the product in such a
deterministic (random) time-space was
$$
[x',x'']^{+,T}:=[s',s'']^{\tau ''}+c^2\left[\tau ',\tau ''\right].
$$
Parallel we used the assumption
that the dimension $n$ is equal to $4$. A particle is a random function $x: I_x \rightarrow S$ satisfying two conditions:
\begin{itemize}
\item the set $I_x\subset T^+$ is an interval
\item $ [x(\tau),x(\tau)]^{\tau}<0 \mbox{ if } \tau\in I_x. $
\end{itemize}
The particle lives on the
interval $I_x$, born at the moment $\inf I_x$ and dies at the moment $\sup I_x$. Since all time-sections of a time-space model is a normed space
of dimension $n$ the Borel sets of the time-sections are independent from time. This means that we could consider the physical properties of
a particle as a trajectory of a stochastic process. A particle is ``realistic" if it holds the ``known laws of physics" and ``idealistic" otherwise.
First we
introduced an inner metric $\delta_{K(\tau)}$ on the space at the moment $\tau$.
\begin{defi}
Let $X(\tau):T\rightarrow \tau K(\tau)$ be a
continuously differentiable (by the time) trajectory of the random function $\left(x(\tau)\mbox{ , }\tau\in I_x\right)$. We say that the
particle $x(\tau)$ is \emph{realistic in its position} if for every $\tau\in I_x$ the random variable $\delta_{K
(\tau)}\left(X(\tau),x(\tau)\right)$ has normal distribution on $\tau K(\tau)$. In other words, the stochastic process $\left(\delta_{K
(\tau)}\left(X(\tau),x(\tau)\right)\mbox{ , }\tau\in I_x\right)$ is a stationary Gaussian process with respect to a given continuously
differentiable function $X(\tau)$. We call the function $X(\tau)$ the \emph{world-line} of the particle $x(\tau)$.
\end{defi}
We note that the
concept of "is realistic in its position" is independent of the choice of $\delta_{K (\tau)}$. As a refinement of this concept we defined another one, which can be considered as a generalization of the principle on the maximality of the speed of the light.
\begin{defi}
We say that a
particle \emph{is realistic in its speed} if it is realistic in its position and the derivatives of its world-line  $X(\tau)$ are time-like
vectors.
\end{defi}
For such two particles $x',x''$ which are realistic in their position we can define a momentary distance by the equality:
$$
\delta(x'(\tau),x''(\tau))=\|X'(\tau)-X''(\tau)\|^{\tau}=\sqrt{[X'(\tau)-X''(\tau),X'(\tau)-X''(\tau)]^{+,T}}.
$$
We could say that two
particles $x'$ and $x''$ agree if the expected value of their distances is equal to zero. Let $I=I_{x'}\cap I_{x''}$ be the common part of
their domains. The required equality is: $$ E(\delta_{K(\tau)}(x'(\tau),x''(\tau)))=\int\limits_{I}\delta_{K(\tau)}(x'(\tau),x''(\tau))\mathrm{
d }\tau= $$ $$ =\int\limits_{I}\|X'(\tau)-X''(\tau)\|^{\tau}\mathrm{ d }\tau=0. $$ In a deterministic time-space we have a function $K(\tau)$,
and we have more possibilities to define orthogonality at a moment $\tau$. We fix a concept of orthogonality and consider it in every normed
space. In the case when the norm is induced by the Euclidean inner product this method should give the same result as the usual concept of
orthogonality. The most natural choice is the concept of Birkhoff orthogonality (see in \cite{gho1}). Using it, in every normed space we can
consider an Auerbach basis (see in \cite{gho1}) which plays the role of a basic coordinate frame. We can determine the coordinates of the
points with respect to this basis. We say that a frame is \emph{at rest with respect to the absolute time} if its origin (as a particle) is at
rest with respect to the absolute time $\tau$ and the unit vectors of its axes are at rest with respect to a fixed Euclidean orthogonal basis of
$S$. In $S$ we fix a Euclidean orthonormal basis and give the coordinates of a point (vector) of $S$ with respect to this basis. We get curves
in $S$ parameterized by the time $\tau$. We define the concept of a frame as follows.

\begin{defi}
The system $\{f_1(\tau),f_2(\tau),f_3(\tau), o(\tau)\}\in (S,\|\cdot\|^{+\tau})\times \tau K(\tau)$ is a \emph{frame}, if
\begin{itemize}
\item
$o(\tau)$ is a particle realistic in its speed, with such a world-line
$$
O(\tau):T\rightarrow \tau K(\tau)
$$
which does
not intersect the absolute time axis $T$,
\item
the functions
$$
f_i(\tau):T\rightarrow \cup\left\{(S,\|\cdot\|^\tau) \mbox{ , } \tau\in
T\right\}
$$
are continuously differentiable, for all fixed $\tau$, \item the system $\{f_1(\tau), f_2(\tau), f_3(\tau)\}$ is an Auerbach basis
with origin $O(\tau)$ in the space $(S,\|\cdot\|^\tau)$.
\end{itemize}
\end{defi}

Note, that for a good model we have to guarantee that Einstein's convention on the equivalence of inertial frames remains valid for us.
However at this time we have no possibility to give the concepts of "frame at rest" and the concept of "frame which moves at a constant velocity with
respect to another one". The reason is that when we changed the norm of the space by the function $K(\tau)$ we concentrated only on the change of the shape of the unit ball and did not use any correspondence between the points of the two unit balls. Obviously, in a concrete computation we
should proceed in the opposite direction, first we should give a correspondence between the points of the old unit ball and the new one and this implies the change of the norm. To this purpose we may define a homotopic mapping $\mathbf{ K }$ which describes the deformation of the norm.
\begin{defi}
Consider a homotopic mapping $ \mathbf{ K }\left(x,\tau\right): (S,\|\cdot\|_E)\times T \rightarrow (S,\|\cdot\|_E) $ holding the assumptions:
\begin{itemize}
\item $\mathbf{ K }\left(x,\tau\right)$ is homogeneous in its first variable and continuously differentiable in its second one,
\item $ \mathbf{ K }\left(\{e_1,e_2,e_3\},\tau\right)$ is an Auerbach basis of $\left(S,\|\cdot\|^{\tau}\right)$ for every $\tau$, \item $
\mathbf{ K }\left(B_E,\tau\right)=K(\tau) $.
\end{itemize}
Then we say that the function $\mathbf{ K }\left(x,\tau\right)$ is the
\emph{shape-function} of the time-space.
\end{defi}

The mapping $\mathbf{ K }\left(x,\tau\right)$ determines the changes at all levels. For example, we can consider a frame is ``at rest" if its
change arises from this globally determined change, and ``moves with constant velocity" if its origin has this property and the directions of
its axes are ``at rest". Precisely, we say, that

\begin{defi} The frame $\{f_1(\tau),f_2(\tau),f_3(\tau),o(\tau)\}$ \emph{ moves at a constant velocity with respect to the time-space}  if for
every pairs $\tau$, $\tau'$ in $T^+$ we have
$$
f_i(\tau )=\mathbf{ K }\left(f_i(\tau'),\tau \right) \mbox{ for all } i \mbox{ with } 1\leq i
\leq 3
$$
and  there are two vectors $O=o_1e_1+o_2e_2+o_3e_3\in S$ and  $v=v_1e_1+v_2e_2+ v_3e_3 \in S$ such that  for all values of $\tau$ we have
$$
O(\tau)=\mathbf{K}(O,\tau)+\tau \mathbf{K}(v,\tau).
$$
A frame is \emph{at rest with respect to the time-space} if the vector $v$ is the zero
vector of $S$.
\end{defi}
Consider the derivative of the above equality by $\tau$. We get that
$$ \dot{O}(\tau)=\frac{\partial
\mathbf{K}(O,\tau)}{\partial \tau}+ \mathbf{K}(v,\tau)+ \tau \frac{\partial \mathbf{K}(v,\tau)}{\partial \tau},
$$
showing that for such a homotopic mapping, which is constant in the time $O(\tau)$, is a line with direction vector $v$ through the origin of the time space. Similarly, in the case when $v$ is the zero vector it is a vertical (parallel to $T$) line-segment through $O$.

We can re-define the concept of time-axes, too.

\begin{defi} The \emph{time-axis} of the time-space model is the world-line $O(\tau)$ of such a particle which moves at a constant velocity with respect to the time-space and starts from the origin. More precisely, for the world-line $\left(O(\tau),\tau\right)$ we have
$\mathbf{K}(O,\tau)=0$ and hence with a given vector $v\in S$,
$$
O(\tau)=\tau \mathbf{K}(v,\tau).
$$
\end{defi}

\begin{remark} Note that if the shape-function is linear in its first variable then all sections defined by $\tau=\mbox{const.}$ are Euclidean
spaces. This is the case when the shape-function is of the form: $$ \mathbf{K}(v,\tau)=f(\tau)A(s), $$ where $f$ is a continuously
differentiable function and $A:S\longrightarrow S$ is a linear mapping.
\end{remark}

\section{On the formulas of special relativity theory}

In this section we assume that the shape-function is a two-times continuously differentiable function, so it is a $C^2$-class function.
We need two further axioms to interpret in time-space the usual axioms of special relativity theory. First we assume that:

\begin{axiom} The laws of physics are invariant under transformations between frames. The laws of physics will be the same whether you are
testing them in a frame "at rest", or a frame moving with a constant velocity relative to the "rest" frame. \end{axiom}

\begin{axiom} The speed of light in a vacuum is measured to be the same by all observers in frames. \end{axiom}

These axioms can be transformed into the language of the time-space by the method of Minkowski \cite{minkowski}. For this we use the imaginary
sphere $H_c$ of parameter $c$ introduced in the previous subsection and the group $G_c$ as the set of those isometries of the space which leave
invariant this sphere of parameter $c$. Such an isometry can be interpreted as a coordinate transformation of the time-space which sends the
axis of the absolute time into another time-axis $t'$, and also maps the intersection point of the absolute time-axis with the imaginary sphere
$H_c$ into the intersection point of the new time-axis and $H_c$. An isometry of the time-space is also a homeomorphism thus it maps the
subspace $S$ into a topological hyperplane $S'$ of the embedding normed space. $S'$ is orthogonal to the new time-axis in the sense that its
tangent hyperplane at the origin is orthogonal to $t'$ with respect to the product of the space. Of course the new space-axes are continuously
differentiable curves in $S'$ whose tangents at the origin are orthogonal to each other. Since the absolute time-axis is orthogonal to the
imaginary sphere $H_c$ the new time-axis $t'$ must hold this property, too. Thus the investigations in the previous section are essential from
this point of view. Assuming that the definition of the time-space implies this property we can get some formulas similar to the well-known
formulas of special relativity. We note that the function $\mathbf{K}(v,\tau)$ holds the orthogonality property of vectors of $S$ and by the
equality
$$
[\mathbf{K}(v,\tau),\mathbf{K}(v,\tau)]^\tau=\|v\|_E^2
$$
we can see that the formulas on time-dilatation and length-contraction are
valid, too. Using the well-known notations
$$
\beta = \frac{\|v\|_E}{c}
$$
$$
\gamma = \frac{1}{\sqrt{1 - \beta^2}}
$$
we get the connection between the time $\tau_0$ and $\tau$ of an event measured by two observers, one at rest and the other one moving at a
constant velocity $\|v\|_E$ with respect to the time-space. It is $$ \tau=\gamma \tau_0. $$ Similarly, we can consider a moving rod whose points move at a constant velocity with respect to the time space such that it is always parallel to the velocity vector $\mathbf{K}(v,\tau)$. Then we have
$$ \|v\|_E=\frac{L_0}{T} $$ where $T$ is the time calculated from the length $L_0$ and the velocity vector $v$ by such an observer which moves
with the rod. Another observer can calculate the length $L$ from the measured time $T_0$ and the velocity $v$ by the formula
$$
\|v\|_E=\frac{L}{T_0}.
$$
Using the above formula of dilatation we get the known Fitzgerald contraction of the rod:
$$
L=L_0\sqrt{1-\beta^2}=\frac{L_0}{\gamma}.
$$

\subsection{Lorentz transformation}

Lorentz transformation in time-space is also based on the usual experiment in which we send a ray of light to a mirror in the direction of the unit vector $e$ with distance $d$ from me.

\subsubsection{Deduction of Lorentz transformation in time-space} If we are at rest, we can determine in time space the respective points $A$,
$C$ and $B$ of departure, turn and arrival of the ray of light. $A$ and $B$ are on the absolute time-axis at heights $\tau_A$, and $\tau_B$,
respectively. The position of $C$ is
$$
(\tau_C-\tau_A)\mathbf{K}(ce,\tau_C-\tau_A)+\tau_C
e_4=\frac{\tau_B-\tau_A}{2}\mathbf{K}\left(ce,\frac{\tau_B-\tau_A}{2}\right)+\frac{\tau_B+\tau_A}{2}e_4,
$$
since we know that the light takes
the road back and forth over the same time. We observe that the norm of the space-like component $s_C$ is
$$
\|s_C\|^{\tau_C}=c\frac{\tau_B-\tau_A}{2}
$$
as in the usual case of space-time.

The moving observer synchronized its clock with the observer at rest in the origin, and moves in the direction $v$ with velocity $\|v\|_E$. We
assume that the moving observer also sees the experiment, thus its time-axis corresponding to the vector $v$ meets the world-line of the light at two points $A'$ and $B'$ positioned on the respective curves $AC$ and $CB$. This implies that the respective space-like components of the
world-line of the light and the world-line of the axis are parallel to each other in every minute. Hence we have:
$$ \|v\|_E
\mathbf{K}(e,\tau)=\mathbf{K}(v,\tau).
$$
From this, we get the equality
$$
\tau_{A'}\mathbf{K}(v,\tau_{A'})+\tau_{A'}e_4=(\tau_{A'}-\tau_A)\mathbf{K}(ce,\tau_{A'}-\tau_A)+\tau_{A'}e_4.
$$
This implies that
$$
{\tau_{A'}}^2{\|v\|_E}^2-c^2{\tau_{A'}}^2=(\tau_{A'}-\tau_A)^2c^2-c^2{\tau_{A'}}^2
$$
and thus
$$
\tau_{A'}=\frac{c}{c-\|v\|_E}\tau_A.
$$
The proper time $(\tau_{A'})_0$ is
$$
(\tau_{A'})_0=\sqrt{1-\beta^2}\frac{c}{c-\|v\|_E}\tau_A=\tau_A\sqrt{\frac{1+\beta}{1-\beta}}.
$$
Similarly we also get that
$$ (\tau_{B'})_0=\tau_B\sqrt{\frac{1-\beta}{1+\beta}},
$$
and we can determine the new time coordinate of the point $C$ with
respect to the new coordinate system:
$$
(\tau_{C})_0=\frac{(\tau_{A'})_0+(\tau_{B'})_0}{2}=
\frac{1}{2}\left(\tau_A\sqrt{\frac{1+\beta}{1-\beta}}+\tau_B\sqrt{\frac{1-\beta}{1+\beta}}\right).
$$
Since the norm of the space-like component
is
$$
\|s_C\|_E=c\frac{\tau_B-\tau_A}{2},
$$
we get that
$$
\tau_A=\tau_C-\frac{\|s_C\|_E}{c} \mbox{ and } \tau_B=\tau_C+\frac{\|s_C\|_E}{c}
$$
and thus
$$
(\tau_{C})_0=\frac{1}{2}\left(\left(\tau_C-\frac{\|s_C\|_E}{c}\right)\sqrt{\frac{1+\beta}{1-\beta}}+
\left(\tau_C+\frac{\|s_C\|_E}{c}\right)\sqrt{\frac{1-\beta}{1+\beta}}\right)=
$$
$$ =\frac{\tau_C-\frac{\beta\|s_C\|_E}{c}}{\sqrt{1-\beta^2}}=
\frac{\tau_C-\frac{\|v\|_E\|s_C\|_E}{c^2}}{\sqrt{1-\frac{\|v\|_E^2}{c^2}}}=
\frac{\tau_C-\frac{[\mathbf{K}(s_C,\tau_C),\mathbf{K}(v,\tau_C)]^{\tau_C}}{c^2}}{\sqrt{1-\frac{\|v\|_E^2}{c^2}}}.
$$
On the other hand, we also have that the space-like component $((s_C)_0)_S$ of the transformed space-like vector $(s_C)_0$  arises also
from a vector parallel to $e$, thus it is of the form
$$
\mathbf{K}(((s_C)_0)_S,\tau)=\|((s_C)_0)_S\|_E \mathbf{K}(e,\tau).
$$
For the norm of $(s_C)_0$ we know that
$$
\|(s_C)_0\|^{+,T}=c\frac{(\tau_{B'})_0-(\tau_{A'})_0}{2},
$$
hence,
$$
\|(s_C)_0\|^{+,T}=\frac{\|s_C\|_E-\|v\|_E\tau_C}{\sqrt{1-\frac{\|v\|_E^2}{c^2}}}.
$$
If we consider the vector
$$
\widehat{(s_C)_0}=\gamma\left(\mathbf{K}(s_C,\tau_C)-\mathbf{K}(v,\tau_C)\tau_C\right)\in S,
$$
we get a norm-preserving, bijective mapping
$\widehat{L}$ from the world-line of the light into $S$ by the definition
$$
\widehat{L}:\mathbf{K}((s_C)_0,(\tau_C)_0)\mapsto
\gamma\left(\mathbf{K}(s_C,\tau_C)-\mathbf{K}(v,\tau_C)\tau_C\right).
$$
The connection between the space-like coordinates of the point with
respect to the two frames now has a more familiar form. Henceforth the Lorentz transformation means for us the correspondence:
\begin{eqnarray*}
s & \mapsto & \widehat{\mathbf{K}(s',\tau')}=\gamma\left(\mathbf{K}(s,\tau)-\mathbf{K}(v,\tau)\tau\right) \\ \tau & \mapsto &
\tau'=\gamma\left(\tau-\frac{[\mathbf{K}(s,\tau),\mathbf{K}(v,\tau)]^{\tau}}{c^2}\right),
\end{eqnarray*}
and the inverse Lorentz transformation the another one
\begin{eqnarray*}
\widehat{\mathbf{K}(s',\tau')} & \mapsto &
\mathbf{K}(s,\tau)=\gamma\left(\mathbf{K}(s',\tau')+\mathbf{K}(v,\tau')\tau'\right) \\ \tau' & \mapsto &
\tau=\gamma\left(\tau'+\frac{[\mathbf{K}(s',\tau'),\mathbf{K}(v,\tau')]^{\tau'}}{c^2}\right).
\end{eqnarray*}

\subsubsection{Consequences of Lorentz transformation}

First, note that we can determine the components of $(s_C)_0$ with respect to the absolute coordinate system, too. Since
$(s_C)_0$ and $\tau\mathbf{K}(v,\tau)+\tau e_4$ are orthogonal to each other we get that
$$
[\mathbf{K}(((s_C)_0)_S,\tau_C),\mathbf{K}(v,\tau_C)]^{\tau_C}=c^2((s_C)_0)_T ,
$$
implying that
$$
((s_C)_0)_T=\frac{\|((s_C)_0)_S\|_E\|v\|_E}{c^2}.
$$
Thus, we get the equality
$$
\|((s_C)_0)_S\|_E^2\left(1-c^2\left(\frac{\|v\|_E}{c^2}\right)^2\right)=
\left(\frac{\|s_C\|_E-\|v\|_E\tau_C}{\sqrt{1-\frac{\|v\|_E^2}{c^2}}}\right)^2,
$$
implying that
$$
\|((s_C)_0)_S\|_E=\frac{\|s_C\|_E-\|v\|_E\tau_C}{\left(1-\frac{\|v\|_E^2}{c^2}\right)}=\gamma ^2\left(\|s_C\|_E-\|v\|_E\tau_C\right)
$$
and
$$
((s_C)_0)_T=\frac{\|((s_C)_0)_S\|_E\|v\|_E}{c^2}=\frac{\|v\|_E\|s_C\|_E-\|v\|_E^2\tau_C}{c^2-\|v\|_E^2}.
$$
We get that
$$ (s_C)_0=\gamma ^2
\left(\|s_C\|_E-\|v\|_E\tau_C\right)\left(\mathbf{K}(e,\tau_C)+\frac{\|v\|_E}{c^2}e_4\right)=
$$
$$ =\gamma
^2\left(\mathbf{K}(s_C,\tau_C)-\mathbf{K}(v,\tau_C)\tau_C\right)+\left(\frac{\gamma}{1-\gamma}\right)^2\left(\|s_C\|_E-\|v\|_E\tau_C\right)e_4.
$$
We can determine the length of this vector in the new coordinate system, too. Since
$$
[(s_C)_0,(s_C)_0]^{+,T}=\left(\|(s_C)_0\|^{+,T}\right)^2=\frac{(\|s_C\|^{\tau_C}-\|v\|_E\tau_C)^2}{1-\frac{\|v\|_E^2}{c^2}}=
$$
$$
=\frac{[s_C,s_C]^{\tau_C}-2\|s_C\|^{\tau_C}\|v\|_E\tau_C +(\|v\|_E\tau_C)^2}{1-\frac{\|v\|_E^2}{c^2}}
$$
and
$$
\left((\tau_{C})_0\right)^2=\frac{(\tau_C)^2-2\tau_C\frac{\|v\|_E\|s_C\|^{\tau_C}}{c^2}+
\frac{\left(\|v\|_E\|s_C\|^{\tau_C}\right)^2}{c^4}}{1-\frac{\|v\|_E^2}{c^2}},
$$
hence the equality
$$
[(s_C)_0,(s_C)_0]^{+,T}-c^2\left((\tau_{C})_0\right)^2=[s_C,s_C]^{\tau_C}-c^2\left(\tau_{C}\right)^2
$$
shows that under the action of the
Lorentz transformation the "norm-squares" of the vectors of the time-space are invariant as in the case of usual space-time.

Finally, we can determine those points of the space whose new time-co\-or\-di\-nates are zero and thus we get a mapping from the subspace $S$ into the time-space. Let $s\in S$ arbitrary and consider the corresponding point $\mathbf{K}(s,\tau)+\tau e_4$ and assume that
$$
0=\tau_0=\gamma
\tau-\gamma \frac{\|v\|_E}{c^2} \|\mathbf{K}(s,\tau)\|^\tau,
$$
hence
$$
\tau=\frac{\|v\|_E \|s\|_E}{c^2}.
$$
Then we get the image of the
coordinate subspace $S$ under the action of that isometry which corresponds to the Lorentz transformation  sending the absolute
time-axis into the time-axis $\tau \mathbf{K}(v,\tau)+\tau e_4$ in question. This set is:
$$
S_0=\left\{ \mathbf{K}\left(s,\frac{\|v\|_E
\|s\|_E}{c^2}\right)+\frac{\|v\|_E \|s\|_E}{c^2} e_4 \quad | \quad s\in S\right\}.
$$

For a boost in an arbitrary direction with velocity $v$,
it is convenient to decompose the spatial vector $s$ into components perpendicular and parallel to $v$:
$$
s=s_1+s_2
$$ so that

$$
[\mathbf{K}(s,\tau),\mathbf{K}(v,\tau)]^\tau = [\mathbf{K}(s_1,\tau),\mathbf{K}(v,\tau)]^\tau + [\mathbf{K}(s_2,\tau),\mathbf{K}(v,\tau)]^\tau =
$$
$$
=[\mathbf{K}(s_2,\tau),\mathbf{K}(v,\tau)]^\tau.
$$

Then, only time and the component $\mathbf{K}(s_2,\tau)$ in the direction of
$\mathbf{K}(v,\tau)$;
\begin{eqnarray*}
\tau' & = &\gamma \left(\tau - \frac{[\mathbf{K}(s,\tau),\mathbf{K}(v,\tau)]^\tau}{c^{2}} \right) \\
\widehat{\mathbf{K}(s',\tau')} & = & \mathbf{K}(s_1,\tau)+ \gamma (\mathbf{K}(s_2,\tau)-\mathbf{K}(v,\tau)\tau )
\end{eqnarray*}
are "distorted"
by the Lorentz factor $\gamma$. The second equality can be written also in the form:
$$
\widehat{s'}=\mathbf{K}(s,\tau)+\left(\frac{\gamma-1}{\|v\|_E^2}[\mathbf{K}(s,\tau),\mathbf{K}(v,\tau)]^\tau-\gamma
\tau\right)\mathbf{K}(v,\tau).
$$
\begin{remark}
If we have two time-axes $\tau \mathbf{K}(v',\tau)+\tau e_4$ and $\tau
\mathbf{K}(v'',\tau)+\tau e_4$ then there are two subgroups of the corresponding Lorentz transformations mapping the absolute time-axis onto
another time-axes, respectively. These two subgroups are also subgroups of $G_c$. Their elements can be paired on the base of their action on
$S$. The pairs of these isometries define a new isometry of the space (and its inverse) in a natural way, with the composition of one of them and the inverse of the other. Omitting the absolute time-axis from the space (as we suggest earlier) the invariance of the product on the remaining space and also the physical axioms of special relativity can remain in effect.
\end{remark}

\subsubsection{Addition of velocities} If
$\mathbf{K}(u,\tau)$ and $\mathbf{K}(v,\tau')$ are two velocity vectors then using the formula for inverse Lorentz transformation of the
corresponding differentials we get that
$$
\mathrm{d}\tau = \gamma \left(\mathrm{d}\tau' +
\frac{[\mathbf{K}(\mathrm{d}\widehat{s'},\mathrm{d}\tau'),\mathbf{K}(v,\tau')]^{\tau'}}{c^{2}} \right)
$$ and
$$
\mathbf{K}(\mathrm{d}s,\mathrm{d}\tau)=\mathbf{K}(\mathrm{d}\widehat{s'},\mathrm{d}\tau')+\left(\frac{1-\gamma}{\|v\|_E^2}
[\mathbf{K}(\mathrm{d}\widehat{s'},\mathrm{d}\tau'),\mathbf{K}(v,\tau')]^{\tau'}+\gamma \mathrm{d}\tau'\right)\mathbf{K}(v,\tau').
$$
Thus
$$
\mathbf{K}(u,\tau)=\frac{\mathbf{K}(\mathrm{d}s,\mathrm{d}\tau)}{\mathrm{d}\tau}=
$$
$$
=\frac{\mathbf{K}(\mathrm{d}\widehat{s'},\mathrm{d}\tau')+\left(\frac{1-\gamma}{\|v\|_E^2}
[\mathbf{K}(\mathrm{d}\widehat{s'},\mathrm{d}\tau'),\mathbf{K}(v,\tau')]^{\tau'}+\gamma \mathrm{d}\tau'\right)\mathbf{K}(v,\tau')}{\gamma
\left(\mathrm{d}\tau' + \frac{[\mathbf{K}(\mathrm{d}\widehat{s'},\mathrm{d}\tau'),\mathbf{K}(v,\tau')]^{\tau'}}{c^{2}} \right)}=
$$
$$
=\frac{\left(\mathbf{K}(v,\tau')+\frac{1}{\gamma}\frac{\mathbf{K}(\mathrm{d}\widehat{s'},\mathrm{d}\tau')}{\mathrm{d}\tau'}+\frac{1+\gamma}{\gamma
c^2}\left[\frac{\mathbf{K}(\mathrm{d}\widehat{s'},
\mathrm{d}\tau')}{\mathrm{d}\tau'},\mathbf{K}(v,\tau')\right]^{\tau'}\mathbf{K}(v,\tau')\right)}
{1+\frac{\left[\frac{\mathbf{K}(\mathrm{d}\widehat{s'},\mathrm{d}\tau')}{\mathrm{d}\tau'},\mathbf{K}(v,\tau')\right]^{\tau'}}{c^{2}}}
$$
$$
=\frac{\left(\mathbf{K}(v,\tau')+ \frac{1}{\gamma}\mathbf{K}(u',\mathrm{d}\tau')+ \frac{1+\gamma}{\gamma
c^2}[\mathbf{K}(u',\mathrm{d}\tau'),\mathbf{K}(v,\tau')]^{\tau'}\mathbf{K}(v,\tau')\right)}
{1+\frac{[\mathbf{K}(u',\mathrm{d}\tau'),\mathbf{K}(v,\tau')]^{\tau'}}{c^{2}}}.
$$

\subsection{Acceleration, momentum and energy}

Our starting point is \emph{ the velocity vector (or four-velocity)}. The absolute time coordinate is $\tau$, this defines a world line of form
$S(\tau)=\mathbf{K}(s(\tau),\tau)+\tau e_4$. Its proper time is $\tau_0=\frac{\tau}{\gamma}=\tau\sqrt{1-\frac{\|v\|_E^2}{c^2}}$, where $v$ is
the velocity vector of the moving frame. By definition
$$
V(\tau):=\frac{\mathrm{d}S(\tau)}{\mathrm{d}\tau_0}=\gamma\left(\frac{\mathrm{d}(\mathbf{K}(s(\tau),\tau))}{\mathrm{d}\tau}+e_4\right).
$$
If the shape-function is a linear mapping then
$\frac{\mathrm{d}(\mathbf{K}(s(\tau),\tau))}{\mathrm{d}\tau}=\mathbf{K}(\dot{s}(\tau),1):=\mathbf{K}(v(\tau),1)$ and we also have
$$
[V(\tau),V(\tau)]^{+,T}=\gamma^2\left([\mathbf{K}(v(\tau),1),\mathbf{K}(v(\tau),1)]^{1}-c^2\right)=-c^2.
$$
The \emph{acceleration} is defined
as the change in four-velocity over the particle's proper time. Hence now the velocity of the particle is also a function of $\tau$ as without
$\gamma $ we have the function $\gamma(\tau)$. The definition is:
$$
A(\tau):=\frac{\mathrm{d}V}{\mathrm{d}\tau_0}=\gamma
(\tau)\frac{\mathrm{d}V}{\mathrm{d}\tau}=
$$
$$
=\gamma^2(\tau)\frac{\mathrm{d}^2 \mathbf{K}(s(\tau),\tau)}{\mathrm{d}\tau^2}+ \gamma(\tau)\gamma'(\tau)
\frac{\mathrm{d}(\mathbf{K}(s(\tau),\tau))}{\mathrm{d}\tau}+\gamma(\tau)\gamma'(\tau)e_4,
$$
where with the notation $a(\tau)=v'(\tau)=s''(\tau)$,
$$ \gamma'(\tau)=\left(\frac{1}{\sqrt{1-\frac{\|v(\tau)\|_E^2}{c^2}}}\right)'=
\left(\frac{1}{\sqrt{1-\frac{\left[\mathbf{K}(v(\tau),1),\mathbf{K}(v(\tau),1)\right]^{1}}{c^2}}}\right)'=
$$
$$
=\frac{\left[\frac{\mathrm{d}(\mathbf{K}(v(\tau),1)}{\mathrm{d}\tau},\mathbf{K}(v(\tau),1)\right]^{1}}
{c^2\left(1-\frac{\left[\mathbf{K}(v(\tau),1),\mathbf{K}(v(\tau),1)\right]^{1}}{c^2}\right)^{\frac{3}{2}}}=
\frac{\left[\frac{\mathrm{d}(\mathbf{K}(v(\tau),1)}{\mathrm{d}\tau},\mathbf{K}(v(\tau),1)\right]^{1}}{c^2}\gamma^3(\tau),
$$
In the case of a linear shape-function it has the form
$$
A(\tau)=\gamma^2(\tau)\mathbf{K}(a(\tau),0)+
\gamma(\tau)\gamma'(\tau)\mathbf{K}(v(\tau),1))+\gamma(\tau)\gamma'(\tau)e_4.
$$
Since in this case $[V(\tau),V(\tau)]^{+,T}=-c^2$, we have
$$
[A(\tau),V(\tau)]^{T,+}=\gamma^3(\tau)\left(\left[\mathbf{K}(a(\tau),0),\mathbf{K}(v(\tau),1)\right]^1+\right.
$$
$$
\left.+\gamma^2(\tau)\frac{\left[\mathbf{K}(a(\tau),0),\mathbf{K}(v(\tau),1)\right]^{1}}{c^2} \|v(\tau)\|_E^2-
\gamma^2(\tau)\left[\mathbf{K}(a(\tau),0),\mathbf{K}(v(\tau),1)\right]^{1}\right)=
$$
$$
=\gamma^3(\tau)\left(\left[\mathbf{K}(a(\tau),0),\mathbf{K}(v(\tau),1)\right]^1- \left[\mathbf{K}(a(\tau),0),\mathbf{K}(v(\tau),1)\right]^{1}\right)=0.
$$
By Theorem 2 on the derivative of the product
(corresponding to smooth and strictly convex norms) we also get this result, in fact we have
$$
0=\frac{\mathrm{d}[V(\tau),V(\tau)]^{+,T}}{\mathrm{d}\tau}=
2\left[\frac{\mathrm{d}V}{\mathrm{d}\tau},V\right]^{+,T}+\frac{\partial[V(\tau),V(\tau)]^\tau}{\partial{\tau}}(1)\cdot
0=
$$
$$
=\frac{2}{\gamma}[A(\tau),V(\tau)]^{+,T}.
$$

Also in the case of a linear shape-function the \emph{momentum} is $$ P=m_0 V=\gamma m_0\left(\mathbf{K}(v(\tau),\tau)+ e_4\right) $$ where
$m_0$ is the invariant mass. We also have that $$ [P,P]^{+,T}=\gamma^2 m_0^2(\|v\|_E^2-c^2)=(m_0c)^2. $$ Similarly the \emph{ force } is $$
F=\frac{\mathrm{d}P}{\mathrm{d}\tau}= m_0\gamma^2(\tau)\mathbf{K}(a(\tau),\tau)+
\gamma(\tau)\gamma'(\tau)\mathbf{K}(v(\tau),\tau))+\gamma(\tau)\gamma'(\tau)e_4, $$ and thus it holds that $$ [F,V]^{+,T}=0. $$

\section{General relativity theory}

In time-space there is a way to describe and visualize certain spaces which are solutions of Einstein's field equations (briefly Einstein's
equation). The first method is when we embed into an at least four-dimensional time-space a four-dimensional manifold whose inner metric is a
solution of Einstein's equation. Our basic references here are the books \cite{eddington} and \cite{griffiths}.

\subsection{Metrics embedded into a time-space}

\subsubsection{Minkowski-Lorentz metric}

The simplest example of a Lorentz manifold is the \emph{flat-space metric} which can be given as $\mathbb{R}^4$ with coordinates $(t,x,y,z)$ and
the metric function:
$$
\mathrm{d}s^2 = -c^2 \mathrm{d}t^2 + \mathrm{d}x^2 + \mathrm{d}y^2 + \mathrm{d}z^2.
$$
In the above coordinates, the
matrix representation is
$$
\eta = \left(\begin{array}{cccc}-1&0&0&0\\0&1&0&0\\0&0&1&0\\0&0&0&1\end{array}\right)
$$
In spherical coordinates
$(t,r,\theta,\phi)$, the flat space metric takes the form
$$
\mathrm{d}s^2 = -c^2 \mathrm{d}t^2 + \mathrm{d}r^2 + r^2 \mathrm{d}\Omega^2.
$$
Here $f(r)\equiv 0$, $g=\mathrm {id }$ and $\tau=t$ implying that $\mathbf{K}\left(v,\tau\right)=v$ and the hypersurface is the light-cone
defined by $\tau=\|v\|_E$. It can be considered also in a $5$-dimensional time-space with shape-function $\mathbf{K}\left(v,\tau\right)=v$ as
the metric of a $4$-dimensional subspace through the absolute time-axis. By the equivalence of time axes in a usual space-time it can be
considered as an arbitrary $4$-dimensional subspace distinct to the $4$-dimensional subspace of space-like vectors, too.

\subsubsection{The de Sitter and the anti-de Sitter metrics}

The \emph{de Sitter space} is the space defined on the de Sitter sphere of a Min\-kows\-ki space of one higher dimension. Usually the metric can be considered as the restriction of the Minkowski metric
$$
\mathrm{d}s^2 = -c^2 \mathrm{d}t^2 + \mathrm{d}x_1^2 + \mathrm{d}x_2^2 +
\mathrm{d}x_3^2+\mathrm{d}x_4^2
$$
to the sphere $-x_0^2+x_1^2+x_2^2+x_3^2+x_4^2=\alpha^2=\frac{3}{\Lambda}$, where $\Lambda $ is the
cosmological constant (see e.g. in \cite{griffiths}). Using also our constant $c$ this latter equation can be rewritten as
$$
-ct^2+(x'_1)^2+(x'_2)^2+(x'_3)^2+(x'_4)^2=1 \mbox{ where } x_0=t \, , \, \frac{1}{\alpha}=c \mbox{ and } x'_i=\frac{1}{\alpha}x_i.
$$
This shows
that in the $5$-dimensional time-space with shape-function $\mathbf{K}\left(v,\tau\right)=v$ it is the hyperboloid with one sheet with circular
symmetry about the absolute time-axis.

The \emph{anti-de Sitter space} is the hyperbolic analogue of the elliptic de Sitter space. The Minkowski space of one higher dimension can be
restricted to the so called \emph{anti-de Sitter sphere} (also called in our terminology as imaginary sphere) defined by the equality
$-x_0^2+x_1^2+x_2^2+x_3^2=-\alpha^2$. The shape function again is $\mathbf{K}\left(v,\tau\right)= v$ and the corresponding $4$-submanifold is
the hyperboloid of two sheets with hyperplane symmetry with respect to the $4$-subspace $S$ of space-time vectors.

\subsubsection{Friedmann-Lema\^{\i}tre-Robertson-Walker metrics}

A standard metric form of the Friedmann-Lema\^{\i}tre-Robertson-Walker metrics (F-L-R-W) family of space-times can be obtained by using
suitable coordinate parameterizations of the 3-spaces of constant curvature. One of its forms is
$$
\mathrm{d}s^2=-\mathrm{d}t^2+\frac{R^2(t)}{1+\frac{1}{4}k(x^2+y^2+z^2)}\left(\mathrm{d}x^2+\mathrm{d}y^2 + \mathrm{d}z^2\right)
$$
where
$k\in\{-1,0,1\}$ is fixed. By the parametrization $\tau=t$ this metric is the metric of a time-space with shape-function $
\mathbf{K}\left(v,\tau\right)$. Observe that
$$
\|v\|_E^2=\left[\mathbf{K}\left(v,\tau\right),\mathbf{K}\left(v,\tau\right)\right]^\tau=
\frac{R^2(\tau)}{1+\frac{1}{4}k\|v\|_E^2}\|\mathbf{K}\left(v,\tau\right)\|_E^2.
$$
Note that we can choose the constant $k$ also as a function
of the absolute time $\tau$ giving a deterministic time-space with more generality. Hence the shape-function is $$
\mathbf{K}\left(v,\tau\right)= \frac{\sqrt{1+\frac{1}{4}k(\tau)\|v\|_E^2}}{R(\tau)}v. $$

\subsection{Three-dimensional visualization of a metric in a four-time-space}

The second method is when we consider a four-dimensional time-space and a three-dimensional sub-manifold in it with the property that the metric
of the time-space at the points of the sub-manifold corresponds to the given one. This method gives a good visualization of the solution
in such a case when the examined metric has some special property e.g. there is no dependence on time or (and) the metric has a spherical symmetry.
The examples of this section are also semi-Riemannian manifolds. We consider now such solutions which have the form: $$ \mathrm{d}s^2 =
-(1-f(r)) c^2 \mathrm{d}t^2 + \frac{1}{1-f(r)} \mathrm{d}r^2 + r^2(\mathrm{d}\theta^2 + \sin^2\theta\mathrm{d}\phi^2) $$ where $$
\mathrm{d}\Omega^2 := \mathrm{d}\theta^2 + \sin^2\theta\mathrm{d}\phi^2 $$ is the standard metric on the 2-sphere. Thus we have to search a
shape function $\mathbf{K}\left(v,\tau\right)$ of the embedding space and a sub-manifold of it on which the Minkowski-metric gives the required
one. If the metric is isotropic we have a chance to give it by isotropic coordinates. We substitute the function $r=g(r^\star)$ into this in place of the parameter $r$, and solve the differential equation: $$ f(g(r^\star))=1-\left(\frac{r^\star g'(r^\star)}{g(r^\star)}\right)^2 $$ for the unknown
function $g(r^\star)$. Then we get the metric in the isotropic form $$ \mathrm{d}s^2 = -\left(\frac{r^\star g'(r^\star)}{g(r^\star)}\right)^2
c^2 \mathrm{d}t^2 + \frac{g^2(r^\star)}{{r^\star}^2} \left(\mathrm{d}{r^\star}^2 + {r^\star}^2(\mathrm{d}\theta^2 + \sin^2\theta
\mathrm{d}\phi^2)\right). $$ For isotropic rectangular coordinates $x=r^\star \sin \theta \cos \phi$, $y=r^\star \sin \theta \sin \phi$ and
$z=r^\star \cos \theta $, the metric becomes $$ \mathrm{d}s^2 = -\left(\frac{r^\star g'(r^\star)}{g(r^\star)}\right)^2 c^2 \mathrm{d}t^2 +
\frac{g^2(r^\star)}{{r^\star}^2} \left(\mathrm{d}x^2 + \mathrm{d}y^2 + \mathrm{d}z^2\right), $$ where $r^\star=\sqrt{x^2+y^2+z^2}$. From this,
substituting $ds^2=0$ and rearranging the equality, we get the velocity of light which is equal to the quantity $$
\sqrt{\frac{\mathrm{d}x^2}{\mathrm{d}t^2} + \frac{\mathrm{d}y^2}{\mathrm{d}t^2} + \frac{\mathrm{d}z^2}{\mathrm{d}t^2}}=\frac{{r^\star}^2
g'(r^\star)}{g^2(r^\star)}c. $$ It is independent of the direction and varies with only the radial distance $r^\star$ (from the point mass at
the origin of the coordinates). At the points of the hypersurface $t=r^\star=\sqrt{x^2+y^2+z^2}$ a metric can be parameterized by time: $$
\mathrm{d}s^2 = -\left(\frac{t g'(t)}{g(t)}\right)^2 c^2 \mathrm{d}t^2 + \frac{g^2(t)}{{t}^2} \left(\mathrm{d}x^2 + \mathrm{d}y^2 +
\mathrm{d}z^2\right), $$ and from the equation $$ \frac{t g'(t)}{g(t)}\mathrm{d}t=\mathrm{d}\tau $$ we can give a re-scale of time by the
parametrization $$ \tau :=\int t\frac{g'(t)}{g(t)}\mathrm{d}t=t\ln(g(t))-\int \ln(g(t))\mathrm{d}t. $$ From this equation we determine the
inverse function $\hat{g}$ for which $t=\hat{g}(\tau)$. Since $\hat{g}(\tau)=t=r^\star=\sqrt{x^2+y^2+z^2}$, we also have that the examined set of points of space-time is a hypersurface defined by the equality: $$ \tau=\left(t\ln(g(t))-\int \ln(g(t))dt\right){\sqrt{x^2+y^2+z^2}}. $$
This implies a new form of the metric at the points of this hypersurface: $$ \mathrm{d}s^2 = -c^2 \mathrm{d}\tau^2 +
\frac{g^2(\hat{g}(\tau))}{{\hat{g}(\tau)}^2} \left(\mathrm{d}x^2 + \mathrm{d}y^2 + \mathrm{d}z^2\right). $$ The corresponding inner product has
the matrix form: $$ \left(
  \begin{array}{cccc}
    -c^2 & 0 & 0 & 0 \\
    0 & \frac{g^2(\hat{g}(\tau))}{{\hat{g}(\tau)}^2} & 0 & 0 \\
    0 & 0 & \frac{g^2(\hat{g}(\tau))}{{\hat{g}(\tau)}^2} & 0 \\
    0 & 0 & 0 & \frac{g^2(\hat{g}(\tau))}{{\hat{g}(\tau)}^2}\\
  \end{array}
\right) $$ and hence the Euclidean lengths of the vectors of the space depend only on the absolute moment $\tau$. Thus we can visualize the
examined metric as a metric at the points of the hypersurface $$ \tau=\left(t\ln(g(t))-\int \ln(g(t))\mathrm{d}t\right)\|v\|_E $$ of certain
time-space. We note that this is not the inner metric of the examined surface of dimension $3$ which can be considered as the metric of a
three-dimensional space-time. To determine the shape-function observe that $$
\|v\|_E^2=\left[\mathbf{K}\left(v,\tau\right),\mathbf{K}\left(v,\tau\right)\right]^\tau=
\frac{g^2(\hat{g}(\tau))}{{\hat{g}(\tau)}^2}\|\mathbf{K}\left(v,\tau\right)\|_E^2 $$ from which we get that $$
\mathbf{K}\left(v,\tau\right)=\frac{{\hat{g}(\tau)}}{g(\hat{g}(\tau))}v. $$

We now give some examples.

\subsubsection{Schwarzschild metric}

Besides the flat space metric the most important metric in general relativity is the \emph{Schwarzschild metric} which can be given in the set
of local polar-coordinates $(t,r,\varphi,\theta )$ by $$ \mathrm{d}s^{2} = -\left(1 - \frac{2GM}{c^2r} \right) c^2 \mathrm{d}t^2 + \left(1 -
\frac{2GM}{c^2r} \right)^{-1} \mathrm{d}r^2 + r^2 \mathrm{d}\Omega^2 $$ where, again, $\mathrm{d}\Omega^2$ is the standard metric on the
2-sphere. Here $G$ is the \emph{gravitational constant} and $M$ is a constant with the dimension of mass. The function $f$ is $$
f(r)=\frac{2GM}{c^2r}:=\frac{r_s}{r} \mbox{ with constant } r_s=\frac{2GM}{c^2}. $$ The differential equation on $g$ is $$
\frac{r_s}{g(r^\star)}=1-\left(\frac{r^\star g'(r^\star)}{g(r^\star)}\right)^2 $$ with the solution $$ g(r^\star)= \frac{r_s}{4}c_1r^\star
{\left( 1 + \frac{1}{c_1 r^\star} \right)}^{2}, $$ and if we choose $\frac{4}{r_s}$ as the parameter $c_1$ we get the known (see in
\cite{eddington}) solution $$ g(r^\star)= r^\star {\left( 1 + \frac{r_s}{4 r^\star} \right)}^{2}. $$ For isotropic rectangular coordinates the
metric becomes $$ \mathrm{d}s^2=-\frac{(1-\frac{r_s}{4r^\star})^{2}}{(1+\frac{r_s}{4r^\star})^{2}} \, c^2 {\mathrm{d} t}^2 +
\left(1+\frac{r_s}{4r^\star}\right)^{4}(\mathrm{d}x^2+\mathrm{d}y^2+\mathrm{d}z^2). $$ The equation between $\tau$ and $t$ is $$
\tau=\int\frac{(1-\frac{r_s}{4t})}{(1+\frac{r_s}{4t})}\mathrm{d}t=\int\frac{4t-r_s}{4t+r_s}\mathrm{d}t=t-2r_s\int\frac{1}{4t+r_s}\mathrm{d}t=
t-\frac{r_s}{2}\ln \left(t+\frac{r_s}{4}\right)+C. $$ Of course we can choose $C=0$. Similarly to the known tortoise-coordinates there is no
explicit inverse function of this parametrization which we denote by $\hat{g}(\tau)=t$. The shape-function of the corresponding time-space is $$
\mathbf{K}\left(v,\tau\right)=\frac{{\hat{g}(\tau)}}{g(\hat{g}(\tau))}v=\left( 1 + \frac{r_s}{4 \hat{g}(\tau)} \right)^{-2}v. $$

\subsubsection{Reissner-Nordstr\"om metric}

In spherical coordinates $(t, r, \theta, \phi)$, the line element for the Reissner-Nord\-str\"om metric is $$ \mathrm{d}s^2 = -\left( 1 -
\frac{r_\mathrm{S}}{r} + \frac{r_Q^2}{r^2} \right) c^2\, \mathrm{d}t^2 + \frac{1}{1 - \frac{r_\mathrm{S}}{r} + \frac{r_Q^2}{r^2}}\,
\mathrm{d}r^2 + r^2\, \mathrm{d}\theta^2 + r^2 \sin^2 \theta \mathrm{d}\phi^2, $$ here again $t$ is the time coordinate (measured by a
stationary clock at infinity), $r$ is the radial coordinate, $r_S= 2GM/c^2$ is the Schwarzschild radius of the body, and $r_Q$ is a
characteristic length scale given by $$ r_{Q}^{2} = \frac{Q^2 G}{4\pi\varepsilon_{0} c^4}. $$ Here $1/4\pi\varepsilon_0$ is the Coulomb force
constant. The function $f$ is $$ f(r)=\frac{r_s}{r}-\frac{r_Q^2}{r^2} $$ The differential equation on $g$ is $$
\frac{r_s}{g(r^\star)}-\frac{r_Q^2}{g^2(r^\star)}=1-\left(\frac{r^\star g'(r^\star)}{g(r^\star)}\right)^2 $$ with the solution $$ g(r^\star)=
\sqrt{\frac{r^2_s}{4}-r^2_Q}\frac{c_1}{2}r^\star {\left( 1 + \frac{1}{c_1 {r^\star}} \right)^2}-\sqrt{\frac{r^2_s}{4}-r^2_Q}+\frac{r_s}{2}, $$
if we choose $c_1:=\frac{2}{\sqrt{\frac{r^2_s}{4}-r^2_Q}}$ we get a more simple form: $$ g(r^\star)= r^\star {\left( 1 +
\frac{\sqrt{\frac{r^2_s}{4}-r^2_Q}}{2{r^\star}} \right)^2}-\sqrt{\frac{r^2_s}{4}-r^2_Q}+\frac{r_s}{2}=r^\star \left( 1 +
\frac{\frac{r^2_s}{4}-r^2_Q}{4{r^\star}^2}\right)+\frac{r_s}{2}. $$ For the isotropic rectangular coordinates we have:
$$ \mathrm{d}s^2 =
-\left(\frac{r^\star\left(1 - \frac{\frac{r^2_s}{4}-r^2_Q}{4{r^\star}^2}\right)}{r^\star \left( 1 +
\frac{\frac{r^2_s}{4}-r^2_Q}{4{r^\star}^2}\right) +\frac{r_s}{2}}\right)^2 c^2 \mathrm{d}t^2 +
$$
$$
+\left(\frac{r^\star \left( 1 +
\frac{\frac{r^2_s}{4}-r^2_Q}{4{r^\star}^2}\right) +\frac{r_s}{2}}{{r^\star}}\right)^2(\mathrm{d}x^2+\mathrm{d}y^2+\mathrm{d}z^2).
$$
Our process now leads to the new time parameter $$
\tau=t-\left(\frac{r_s}{4}-\frac{r_Q}{2}\right)\ln\left(\left(t+\frac{r_s}{4}\right)^2-\frac{r_Q^2}{4}\right)-
r_Q\ln\left(t+\frac{r_s}{4}+\frac{r_Q}{2}\right)+C, $$ which, in the case of $C=r_Q=0$, gives back the parametrization of Schwarzschild solution.
The shape-function of the searched time-space can be determined by the corresponding inverse $t=\hat{g}(\tau)$, it is $$
\mathbf{K}\left(v,\tau\right)=\frac{{\hat{g}(\tau)}}{g(\hat{g}(\tau))}v=\frac{\hat{g}(\tau)}{\hat{g}(\tau) \left( 1 +
\frac{\frac{r^2_s}{4}-r^2_Q}{4{\hat{g}(\tau)}^2}\right)+\frac{r_s}{2}}v. $$

Analogously we can compute the time-space visualization of the Schwarz\-schild-de Sitter solution which we now omit.

\subsubsection{Bertotti-Robinson metric}

The Bertotti-Robinson space-time is the only conformally flat solution of the Einstein-Maxwell equalities for a non-null source-free
electromagnetic field. The metric is: $$ \mathrm{d}s^2 =\frac{Q^2}{r^2}\left(- \mathrm{d}t^2 + \mathrm{d}x^2 + \mathrm{d}y^2 +
\mathrm{d}z^2\right), $$ and on the light-cone $t=r$ it has the form $$ \mathrm{d}s^2 =-\frac{Q^2}{t^2}\mathrm{d}t^2 +
\frac{e^2}{t^2}\left(\mathrm{d}x^2 + \mathrm{d}y^2 + \mathrm{d}z^2 \right). $$ By the new time coordinate $$ \tau=Q\ln t \mbox{ or }
t=e^{\frac{\tau}{Q}} $$ using orthogonal space coordinates we get the form $$ \mathrm{d}s^2=-\mathrm{d} \tau^2 +
\frac{Q^2}{e^{\frac{2\tau}{Q}}}\left(\mathrm{d}x^2+\mathrm{d}y^2+\mathrm{d}z^2\right). $$ Thus it can be visualized on the hypersurface
$\tau=e\ln r$ of the time-space with shape-function: $$ \mathbf{K}\left(v,\tau\right):=\frac{e^{\frac{\tau}{Q}}}{Q}v. $$

\subsection{Einstein fields equations}

As we saw in the previous section the direct embedding of a solution of Einstein's equation into a time-space requires non-linear and very
complicated shape-functions. It can be seen also that there are such solutions for which there are no embeddings into a time-space. This
motivates the investigations of the present section. Our build-up follows the one of the clear paper of Prof. Alan Heavens \cite{heavens}, we
would like to thank to him for his downloadable PDF.

\subsubsection{Homogeneous time-space-manifolds and the Equivalence Principle}

We consider now such manifolds whose tangent spaces are four-dimensional time-spaces with given shape-functions. More precisely: \begin{defi}
Let $\mathcal{S}$ be the set of linear mappings $\mathbf{K}(v,\tau):\mathbb{E}^3\times \mathbb{R}\longrightarrow \mathbb{E}^3$ satisfying the
properties of a linear shape-function given in Definition 7. Giving to it the natural topology we say that ${K}$ is \emph{the space of
shape-functions}. If we have a four-dimensional topological manifold $M$ and a smooth ($C^\infty$) mapping $\mathcal{K}:M\longrightarrow
\mathcal{S}$ with the property that at the point $P\in M$ the tangent space is the time-space defined by $\mathbf{K}^P(s,\tau)\in \mathcal{S}$
we say that this pair is a \emph{time-space-manifold}. The time-space manifold is \emph{homogeneous} if the mapping $\mathcal{K}$ is a constant
function. \end{defi} Note that a Lorentzian manifold is such a homogeneous time-space manifold whose shape-function is independent of time
and it is the identity mapping on its space-like components, namely $\mathbf{K}^P(s,\tau)=s$ for all $P$ and for all $\tau $. Its matrix-form
(using the column representation of vectors in time-space) is: $$ \left(
  \begin{array}{cccc}
    1 & 0 & 0 & 0 \\
    0 & 1 & 0 & 0 \\
    0 & 0 & 1 & 0 \\
  \end{array}
\right) $$ Our purpose is to build up the theory of global relativity in  homogeneous time-space-manifolds. We accept the so-called \emph{Strong
Equivalence Principle} of Einstein in the following form: \begin{axiom}(Equivalence Principle) At any point in a homogeneous time-space manifold
it is possible to choose a \emph{locally-inertial frame} in which the laws of physics are the same as the special relativity of the
corresponding time-space. \end{axiom} According to this principle, there is a coordinate-system in which a freely-moving particle moves at a
constant velocity with respect to the time-space $\mathcal{K}(P)=\mathbf{K}^P(s,\tau)=\mathbf{K}(s,\tau)$. It is convenient to write the world
line $$ S(\tau)=\mathbf{K}(s(\tau),\tau)+\tau e_4 $$ parametrically, as a function of the proper time $\tau_0=\frac{\tau}{\gamma(\tau)}$. In
subsection 2.2 we determined the velocity using the time-space parameter $\tau$: $$
V(\tau)=\gamma(\tau)\left(\frac{\mathrm{d}(\mathbf{K}(s(\tau),\tau))}{\mathrm{d}\tau}+e_4\right)=\gamma(\tau)\left(\mathbf{K}(v(\tau),1)+e_4\right).
$$ Taking into consideration again that the shape-function is linear, the acceleration is : $$ A(\tau)=\gamma^2(\tau)\mathbf{K}(a(\tau),0)+
\gamma^4(\tau)\frac{\left[\mathbf{K}(a(\tau),0),\mathbf{K}(v(\tau),1)\right]^{\tau}}{c^2} \mathbf{K}(v(\tau),1)+ $$ $$
+\gamma^4(\tau)\frac{\left[\mathbf{K}(a(\tau),0),\mathbf{K}(v(\tau),1)\right]^{\tau}}{c^2} e_4, $$ giving the differential equation $A(\tau)=0$
for such a particle which moves linearly with respect to this frame.

\subsubsection{Affine connection and the metric on a homogeneous time-space-manifold}

Consider any other coordinate system in which the particle coordinates are $S'(\tau_0)$. Using the chain rule, the defining equation $$
0=A(\tau_0)=\frac{\mathrm{d} V(\tau_0)}{\mathrm{d} \tau_0}=\frac{\mathrm{d}^2 S(\tau_0)}{\mathrm{d}\tau_0^2} $$ becomes $$
0=\frac{\mathrm{d}}{\mathrm{d}\tau_0}\left(\frac{\mathrm{d}{S}}{\mathrm{d}{S'}}\frac{\mathrm{d}S'(\tau_0)}{\mathrm{d}\tau_0}\right)=
\frac{\mathrm{d}{S}}{\mathrm{d}{S'}}\frac{\mathrm{d}^2S'(\tau_0)}{\mathrm{d}\tau_0^2}+
\frac{\mathrm{d}}{\mathrm{d}\tau_0}\left(\frac{\mathrm{d}{S}}{\mathrm{d}{S'}}\right)\frac{\mathrm{d}S'(\tau_0)}{\mathrm{d}\tau_0}= $$ $$
=\frac{\mathrm{d}{S}}{\mathrm{d}{S'}}\frac{\mathrm{d}^2S'(\tau_0)}{\mathrm{d}\tau_0^2}+
\frac{\mathrm{d}^2{S}}{\mathrm{d}{S'}\mathrm{d}{S'}}\frac{\mathrm{d}S'(\tau_0)}{\mathrm{d}\tau_0}\frac{\mathrm{d}S'(\tau_0)}{\mathrm{d}\tau_0},
$$ where $\frac{\mathrm{d}{S}}{\mathrm{d}{S'}}$ means the total derivatives of the mapping of the time-space sending the path $S'(\tau_0)$ into
the specific path $S(\tau_0)$, and the trilinear function $\frac{\mathrm{d}^2{S}}{\mathrm{d}{S'}\mathrm{d}{S'}}$ is the second total derivatives
of the same mapping. (If there is a general smooth transformation between the coordinate-frames, the corresponding derivatives exist.) From
this equality we get the tensor form of the so called \emph{geodesic equation} of homogeneous time-space manifold, it is:
$$
\frac{\mathrm{d}^2S'(\tau_0)}{\mathrm{d}\tau_0^2}+\left(\frac{\mathrm{d}{S'}}{\mathrm{d}{S}}\frac{\mathrm{d}^2{S}}{\mathrm{d}{S'}\mathrm{d}{S'}}\right)
\frac{\mathrm{d}S'(\tau_0)}{\mathrm{d}\tau_0}\frac{\mathrm{d}S'(\tau_0)}{\mathrm{d}\tau_0}=
$$
$$
=\frac{\mathrm{d}^2S'(\tau_0)}{\mathrm{d}\tau_0^2}+
\Gamma(S',S)\frac{\mathrm{d}S'(\tau_0)}{\mathrm{d}\tau_0}\frac{\mathrm{d}S'(\tau_0)}{\mathrm{d}\tau_0}=0.
$$
Here we denote the inverse of the
total derivatives $\frac{\mathrm{d}{S}}{\mathrm{d}{S'}}$ by $\frac{\mathrm{d}{S'}}{\mathrm{d}{S}}$. The name of $\Gamma (S',S)$ is the
\emph{affine connection}.

For uniform labeling we denote by $x^4$ the identity function. Since the shape function is a linear mapping we can represent it as the
multiplication on left by the $3\times 4$ matrix $K=[k_{ij}]=k^i{}_j$. In the rest of this paragraph we apply all conventions of general
relativity. The Greek alphabet is used for space and time components, where indices take values 1,2,3,4 (frequently used letters are
$\mu,\nu,\cdots$) and the Latin alphabet is used for spatial components only, where indices take values 1,2,3 (frequently used letters are $i,
j, ...$) and according to Einstein's convention, when an index variable appears twice in a single term it implies summation of that term
over all the values of the index. The upper indices are indices of coordinates, coefficients or basis vectors.

The mapping $\mathcal{S}:S'(\tau_0)\longrightarrow S(\tau_0)$ sends  $K({x'}^1,{x'}^2,{x'}^3,{x'}^4)^T+{x'}^4e_4$ into the vector
$K(x^1,x^2,x^3,x^4)^T+x^4e_4$. Denote by $\widetilde{K}$ the $4\times 4$ matrix with coefficients: $$ \left(
  \begin{array}{cccc}
    k^1{}_{1} & k^{1}{}_2 & k^{1}{}_3 & k^{1}{}_4 \\
    k^2{}_{1} & k^2{}_{2} & k^2{}_{3} & k^2{}_{4} \\
    k^3{}_{1} & k^3{}_{2} & k^3{}_{3} & k^3{}_{4} \\
    0 & 0 & 0 & 1 \\
  \end{array}
\right), $$ then we get $\mathcal{S}:\widetilde{K}({x'}^1,{x'}^2,{x'}^3,{x'}^4)^T\mapsto \widetilde{K}(x^1,x^2,x^3,x^4)^T$. If the
shape-function $\mathbf{K}$ restricted to the subspace $S$ is a regular linear mapping then we also have $$
\widetilde{K}^{-1}\mathcal{S}\widetilde{K}({x'}^1,{x'}^2,{x'}^3,{x'}^4)^T=(x^1,x^2,x^3,x^4)^T $$ and we have that $$ \left[\frac{\partial
x^\alpha}{\partial {x'}^\mu}\right]=\frac{\mathrm{d} \widetilde{K}^{-1}\mathcal{S}\widetilde{K}}{\mathrm{d} S'}=
\widetilde{K}^{-1}\frac{\mathrm{d} \mathcal{S}}{\mathrm{d} S'}\widetilde{K} \mbox{ and so } \frac{\mathrm{d} \mathcal{S}}{\mathrm{d}
S'}=\widetilde{K}\left[\frac{\partial x^\alpha}{\partial {x'}^\mu}\right]\widetilde{K}^{-1}. $$
Hence
$$
\frac{\mathrm{d}{S'}}{\mathrm{d}{S}}=\widetilde{K}\left[\frac{\partial x^\alpha}{\partial
{x'}^\mu}\right]^{-1}\widetilde{K}^{-1}=\widetilde{K}\left[\frac{\partial {x'}^\mu}{\partial x^\alpha}\right]\widetilde{K}^{-1}   \mbox{ and }
\left[\frac{\mathrm{d}^2{S}}{\mathrm{d}{S'}\mathrm{d}{S'}}\right]^\alpha=
$$
$$
=\widetilde{K}\left[\frac{\partial^2 x^\alpha}{\partial {x'}^\mu\partial
{x'}^\nu}\right]\widetilde{K}^{-1}
$$
implying that the affine connection is: $$ \Gamma(S',S)^{\lambda}{}_{\mu\nu}=\widetilde{K}\frac{\partial
{x'}^\lambda}{\partial x^\alpha}\frac{\partial^2 {x}^\alpha}{\partial {x'}^\mu\partial
{x'}^\nu}\widetilde{K}^{-1}=\widetilde{K}\Gamma^{\lambda}{}_{\mu\nu}\widetilde{K}^{-1}=\widetilde{K}\left\{ \begin{array}{c} \lambda\\ \mu\nu
\end{array} \right\}\widetilde{K}^{-1}. $$ Since $S'(\tau_0)=\widetilde{K}({x'}^1,{x'}^2,{x'}^3,{x'}^4)^T$ thus we also get three equalities,
the first one is: $$
\frac{\mathrm{d}S'(\tau_0)}{\mathrm{d}\tau_0}=\widetilde{K}\left(\frac{\mathrm{d}{x'}^1}{\mathrm{d}\tau_0},\frac{\mathrm{d}{x'}^2}{\mathrm{d}\tau_0},
\frac{\mathrm{d}{x'}^3}{\mathrm{d}\tau_0},\frac{\mathrm{d}{x'}^4}{\mathrm{d}\tau_0}\right)^T=
$$
$$
=\left(k^1{}_{\alpha}\frac{\mathrm{d}{x'}^\alpha}{\mathrm{d}\tau_0},k^2{}_{\alpha}\frac{\mathrm{d}{x'}^\alpha}{\mathrm{d}\tau_0},
k^3{}_{\alpha}\frac{\mathrm{d}{x'}^\alpha}{\mathrm{d}\tau_0}, k^4{}_{\alpha}\frac{\mathrm{d}{x'}^\alpha}{\mathrm{d}\tau_0}\right)^T
=\left[k^\lambda{}_{\alpha}\frac{\mathrm{d}{x'}^\alpha}{\mathrm{d}\tau_0}\right].
$$
The second equality is:
$$
\frac{\mathrm{d}S'(\tau_0)}{\mathrm{d}\tau_0}\frac{\mathrm{d}S'(\tau_0)}{\mathrm{d}\tau_0}=
$$
$$
=\widetilde{K}\left(\frac{\mathrm{d}{x'}^1}{\mathrm{d}\tau_0},\frac{\mathrm{d}{x'}^2}{\mathrm{d}\tau_0},
\frac{\mathrm{d}{x'}^3}{\mathrm{d}\tau_0},\frac{\mathrm{d}{x'}^4}{\mathrm{d}\tau_0}\right)^T
\left(\frac{\mathrm{d}{x'}^1}{\mathrm{d}\tau_0},\frac{\mathrm{d}{x'}^2}{\mathrm{d}\tau_0},
\frac{\mathrm{d}{x'}^3}{\mathrm{d}\tau_0},\frac{\mathrm{d}{x'}^4}{\mathrm{d}\tau_0}\right)\widetilde{K}^T=
$$
$$
=\widetilde{K}\left[\frac{\mathrm{d}{x'}^\mu}{\mathrm{d}\tau_0} \frac{\mathrm{d}{x'}^\nu}{\mathrm{d}\tau_0}\right]\widetilde{K}^T,
$$
and the third one is:
$$
\frac{\mathrm{d}^2S'(\tau_0)}{\mathrm{d}\tau_0^2}=\widetilde{K}\left(\frac{\mathrm{d}^2{x'}^1}{\mathrm{d}\tau_0^2},
\frac{\mathrm{d}^2{x'}^2}{\mathrm{d}\tau_0^2},
\frac{\mathrm{d}^2{x'}^3}{\mathrm{d}\tau_0^2},\frac{\mathrm{d}^2{x'}^4}{\mathrm{d}\tau_0^2}\right)^T=
\left[k^\lambda{}_{\alpha}\frac{\mathrm{d}^2{x'}^\alpha}{\mathrm{d}\tau_0^2}\right].
$$
The geodesic equation now is:
$$
0=\widetilde{K}\left(\frac{\mathrm{d}^2{x'}^1}{\mathrm{d}\tau_0^2}, \frac{\mathrm{d}^2{x'}^2}{\mathrm{d}\tau_0^2},
\frac{\mathrm{d}^2{x'}^3}{\mathrm{d}\tau_0^2},\frac{\mathrm{d}^2{x'}^4}{\mathrm{d}\tau_0^2}\right)^T+
\widetilde{K}\Gamma^{\lambda}{}_{\mu\nu}\widetilde{K}^{-1}\widetilde{K}\left[\frac{\mathrm{d}{x'}^\mu}{\mathrm{d}\tau_0}
\frac{\mathrm{d}{x'}^\nu}{\mathrm{d}\tau_0}\right]\widetilde{K}^T,
$$
or equivalently
$$
0=\left(\frac{\mathrm{d}^2{x'}^1}{\mathrm{d}\tau_0^2},
\frac{\mathrm{d}^2{x'}^2}{\mathrm{d}\tau_0^2},
\frac{\mathrm{d}^2{x'}^3}{\mathrm{d}\tau_0^2},\frac{\mathrm{d}^2{x'}^4}{\mathrm{d}\tau_0^2}\right)^T+
\Gamma^{\lambda}{}_{\mu\nu}\left[\frac{\mathrm{d}{x'}^\mu}{\mathrm{d}\tau_0} \frac{\mathrm{d}{x'}^\nu}{\mathrm{d}\tau_0}\right]\widetilde{K}^T,
$$
implying that
$$
0=
\frac{\mathrm{d}^2{x'}^\lambda}{\mathrm{d}\tau_0^2}+\Gamma^{\lambda}{}_{\mu\nu}\frac{\mathrm{d}{x'}^\mu}{\mathrm{d}\tau_0}k^\nu{}_\zeta
\frac{\mathrm{d}{x'}^\zeta}{\mathrm{d}\tau_0}. $$ Since for the proper time we have the equality $$ -c^2\mathrm{d}\tau_0^2=\mathrm{d}S^T\left(
                                        \begin{array}{cc}
                                          1 & 0\\
                                          0 & -c^2 \\
                                        \end{array}
                                      \right)\mathrm{d}S=\left(\frac{\mathrm{d}S}{\mathrm{d}S'}\mathrm{d}S'\right)^T\eta
                                      \frac{\mathrm{d}S}{\mathrm{d}S'}\mathrm{d}S'=
\mathrm{d}S'^Tg(S',S)\mathrm{d}S'
$$
hence
$$
g(S',S)=\left(\frac{\mathrm{d}S}{\mathrm{d}S'}\right)^T\eta \frac{\mathrm{d}S}{\mathrm{d}S'}.
$$
Let us denote by $[{}_j{}^{i}k]$ the transpose of the matrix $[k^{i}{}_j]$, and let $K^{i}{}_j$ be the elements of the inverse of $\widetilde{K}$. Then, since
$$
g(S',S)=\left(\widetilde{K}^{-1}\right)^T\left[\frac{\partial x^\alpha}{\partial
{x'}^\mu}\right]^T\widetilde{K}^{T}\eta\widetilde{K}\left[\frac{\partial x^\alpha}{\partial {x'}^\mu}\right]\widetilde{K}^{-1} ,
$$
we have
$$
g(S',S)_{\varphi\psi}={}_\varphi{}^\mu{K}\frac{\partial x^\alpha}{\partial {x'}^\mu}{}_\alpha{}^{\delta}k \eta_{\delta,\varepsilon}
k^\varepsilon{}_\beta\frac{\partial x^\beta}{\partial {x'}^\nu}K^{\nu}{}_\psi.
$$
This matrix is the \emph{metric tensor} of the homogeneous
time-space manifold in question. If $\widetilde{K}$ is the unit matrix, then $\mu=\varphi$, $\nu=\psi$, $\alpha=\delta$ and $\beta=\varepsilon$,
implying the known formula
$$
g_{\mu\nu}=\frac{\partial x^\alpha}{\partial {x'}^\mu}\frac{\partial x^\beta}{\partial
{x'}^\nu}\eta_{\alpha\beta}.
$$
Also note that if $\widetilde{K}$ is an orthogonal transformation then we get a simpler form of the metric:
$$
g(S',S)=\widetilde{K}\left[\frac{\partial x^l}{\partial {x'}^i}\right]^T\eta\left[\frac{\partial x^l}{\partial
{x'}^i}\right]\widetilde{K}^{T}.
$$
To determine the connection between the metric and the affine connection, we determine the partial
derivative of the metric.
$$
\frac{\partial g(S',S)}{\partial {x'}^\lambda}=\left(\widetilde{K}^{-1}\right)^T\left[\frac{\partial^2
x^\alpha}{\partial {x'}^\mu\partial {x'}^\lambda}\right]^T\widetilde{K}^{T}\eta\widetilde{K}\left[\frac{\partial x^\beta}{\partial
{x'}^\nu}\right]\widetilde{K}^{-1}+
$$
$$
+\left(\widetilde{K}^{-1}\right)^T\left[\frac{\partial x^\alpha}{\partial
{x'}^\mu}\right]^T\widetilde{K}^{T}\eta\widetilde{K}\left[\frac{\partial^2 x^\beta}{\partial {x'}^\nu\partial
{x'}^\lambda}\right]\widetilde{K}^{-1},
$$
and since
$$ \frac{\partial^2 {x}^\alpha}{\partial {x'}^\mu\partial {x'}^\lambda}=\frac{\partial
{x}^\alpha}{\partial {x'}^\rho}\widetilde{K}^{-1}\Gamma(S',S)^{\rho}{}_{\mu\lambda}\widetilde{K}
$$
we have
$$
\frac{\partial
g(S',S)_{\varphi\psi}}{\partial
{x'}^\lambda}=\Gamma(S',S)^{\rho}{}_{\varphi\lambda}g(S',S)_{\rho\psi}+g(S',S)_{\varphi\rho}\Gamma(S',S)^{\rho}{}_{\lambda\psi}
$$
as in the classical case. Denote by $g(S,S')^{\varphi\rho}$ the inverse of the metric tensor. Then we get the connection:
$$
\Gamma(S',S)^{\sigma}{}_{\lambda\mu}=\frac{1}{2}g(S,S')^{\nu\sigma}\left\{\frac{\partial g(S',S)_{\mu,\nu}}{\partial {x'}^\lambda}+
\frac{\partial g(S',S)_{\lambda,\nu}}{\partial {x'}^\mu}-\frac{\partial g(S',S)_{\mu,\lambda}}{\partial {x'}^\nu}\right\}.
$$

\subsubsection{Covariant derivative, parallel transport and the curvature tensor}

Since we determined the affine connection we can define the
\emph{covariant derivative} of a vector field in the way:
$$
V^\mu{}_{;\lambda}=\frac{\partial V^\mu}{\partial
{x'}^\lambda}+\Gamma(S',S)^{\mu}{}_{\lambda\rho}V^\rho= \frac{\partial V^\mu}{\partial
{x'}^\lambda}+\widetilde{K}\Gamma^{\mu}{}_{\lambda\delta}\widetilde{K}^{-1}V^\delta.
$$
In fact, it converts vectors into tensors on the basis of
the following calculation:
$$
\widetilde{K}\left[\frac{\partial {x'}^\mu}{\partial x^\nu}\right]\left[\frac{\partial x^\rho}{\partial
{x'}^\lambda}\right]\widetilde{K}^{-1}V^\nu{}_{;\rho}=
$$
$$
=\widetilde{K}\left[\frac{\partial {x'}^\mu}{\partial x^\nu}\right]\left[\frac{\partial
x^\rho}{\partial {x'}^\lambda}\right]\widetilde{K}^{-1}\left(\frac{\partial V^\nu}{\partial
{x}^\rho}+\widetilde{K}\Gamma^{\nu}{}_{\rho\delta}\widetilde{K}^{-1}V^\delta\right)=
$$
$$
=\widetilde{K}\left[\frac{\partial {x'}^\mu}{\partial
x^\nu}\right]\left[\frac{\partial x^\rho}{\partial {x'}^\lambda}\right]\widetilde{K}^{-1}\left(\frac{\partial V^\nu}{\partial
{x}^\rho}+\widetilde{K} \frac{\partial {x'}^\nu}{\partial x^\alpha}\frac{\partial^2 {x}^\alpha}{\partial {x'}^\rho\partial
{x'}^\delta}\widetilde{K}^{-1}V^\delta\right)=
$$
$$
=\frac{\partial V'^\mu}{\partial {x'}^\lambda}+\widetilde{K}\frac{\partial
{x'}^\mu}{\partial x^\alpha}\frac{\partial^2 {x}^\alpha}{\partial {x'}^\lambda\partial {x'}^\delta}\widetilde{K}^{-1}{V'}^\delta=\frac{\partial
{V'}^\mu}{\partial {x'}^\lambda}+\widetilde{K}\Gamma^{\mu}{}_{\lambda\delta}\widetilde{K}^{-1}{V'}^\delta={V'}^\mu{}_{;\lambda}.
$$
Note that
the covariant derivative of a co-vector is
$$
V_{\mu;\lambda}=\frac{\partial V_\mu}{\partial
{x'}^\lambda}-\Gamma(S',S)^{\mu}{}_{\lambda\rho}V^\rho,
$$
and the covariant derivative of a tensor has the rule that each upper index adds a
$\Gamma$ term and each lower index subtracts one. For this reason the covariant derivative of the metric tensor (by our calculation above)
vanishes.

Again from the definition of the covariant derivative we get that the \emph{equation of parallel transport} is now:
$$
\frac{\mathrm{d} V^\mu}{\mathrm{d}\tau_0}=-\Gamma(S',S)^{\mu}{}_{\lambda\nu}\frac{\mathrm{d}{x'}^\lambda}{\mathrm{d}\tau_0}V^\nu.
$$
From this it follows that the parallel-transport along a side $\delta {x'}^\beta$ of a small closed parallelogram is
$$
\delta V^\alpha
=-\Gamma(S',S)^{\alpha}{}_{\beta\nu}V^\nu\delta {x'}^\beta
$$
and thus the total change around a small closed parallelogram with sides $\delta
a^\mu$, $\delta b^\nu$ is
$$
\delta V^\alpha=\left(\Gamma(S',S)^{\alpha}{}_{\beta\nu;\rho}V^\nu+\Gamma(S',S)^{\alpha}{}_{\beta\nu}V^\nu{}_{;\rho}-
\Gamma(S',S)^{\alpha}{}_{\rho\nu;\beta}V^\nu-\right.
$$
$$
\left.-\Gamma(S',S)^{\alpha}{}_{\rho\nu}V^\nu{}_{;\beta}\right)\delta a^\beta\delta b^{\rho}
$$
implying that
$$
\delta V^\alpha=R(S',S)^{\alpha}{}_{\sigma\rho\beta}V^{\sigma}\delta a^\beta\delta b^{\rho}.
$$
Here
$R(S',S)^{\alpha}{}_{\sigma\rho\beta}$ is the \emph{Riemann curvature tensor} defined by
$$
R(S',S)^{\alpha}{}_{\sigma\rho\beta}:=\Gamma(S',S)^{\alpha}{}_{\beta\sigma;\rho}-\Gamma(S',S)^{\alpha}{}_{\rho\sigma;\beta}
+\Gamma(S',S)^{\alpha}_{\rho\nu}\Gamma(S',S)^{\nu}_{\sigma\beta}-
$$
$$
-\Gamma(S',S)^{\alpha}_{\beta\nu}\Gamma(S',S)^{\nu}_{\sigma\rho}.
$$
The Ricci Tensor and the scalar curvature are defined by $$ R(S',S)_{\sigma\beta}:=R(S',S)^{\alpha}{}_{\sigma\alpha\beta} \mbox{ and }
R(S',S):=R(S',S)^\sigma{}_\sigma, $$ respectively.

\subsubsection{Einstein's equation}

As we saw in the previous paragraph all of the notions of global relativity can be defined in a time-space-manifold, thus all the equations
between them are well-defined equations. On the other hand, Einstein's equation takes into consideration the facts of physics; hence it contains
parameters which can not be changed. Fortunately, we noted earlier that the covariant derivative of our metric tensor vanishes, too. Thus also
does its inverse vanish, and hence we can write Einstein's equation with \emph{cosmological constant} $\Lambda$, too.
The equation is formally the same as the original one, but it contains a new (undetermined) parameter which is the matrix $\widetilde{K}$ of the
shape-function. It is: $$ R(S',S)^{\mu\nu}-\frac{1}{2}g(S',S)^{\mu\nu}R(S',S)-\Lambda g(S',S)^{\mu\nu}=\frac{8\pi G}{c^4}T^{\mu\nu}, $$ where
the parameter $G$ can be adjusted so that the active and gravitational masses are equal and $T^{\mu\nu}$ is the \emph{energy-momentum tensor}.

\end{document}